\documentclass[aps,prd,reprint,longbibliography,nofootinbib,superscriptaddress,floatfix]{revtex4-2}

\usepackage{slashed}
\usepackage{verbatim}
\usepackage[T1]{fontenc}
\usepackage{mathbbol}
\usepackage[dvipsnames]{xcolor}
\usepackage{orcidlink}
\usepackage{ragged2e}
\usepackage[normalem]{ulem}
\usepackage[english]{babel}
\usepackage{lipsum}
\usepackage{physics}
\usepackage{dcolumn}
\usepackage{tensor}
\usepackage{comment}
\usepackage{placeins}
\usepackage{graphicx,color,overpic,mathtools}
\usepackage{amsthm,amsmath,amssymb,mathrsfs}
\usepackage{braket,bm,bbm,setspace}
\usepackage{booktabs}
\usepackage{cancel}
\usepackage{float}
\usepackage{xargs}

\usepackage{hyperref}
\hypersetup{
    colorlinks=true,
    linkcolor=RoyalBlue,
    citecolor=ForestGreen,
    urlcolor=RoyalBlue}
\definecolor{myred}{RGB}{179, 27, 27}
 
\def\be{\begin{equation}}
\def\ee{\end{equation}}
\def\bea{\begin{eqnarray}}
\def\eea{\end{eqnarray}}
\def\beq{\begin{eqnarray}}
\def\eeq{\end{eqnarray}}

\providecommand{\apj}{Astrophysical Journal}
\providecommand{\apjl}{Astrophysical Journal Letters}

\begin{document}

\title{Dynamical Friction as Environmental Gravitational Self-Force}

\author{Sayak Datta}
\affiliation{Gran Sasso Science Institute (GSSI), I-67100 L'Aquila, Italy}
\affiliation{INFN, Laboratori Nazionali del Gran Sasso, I-67100 Assergi, Italy}

\date{\today}

\begin{abstract}
Dynamical friction (DF) and gravitational-wave radiation reaction are conventionally treated as distinct dissipative mechanisms acting on a compact object inspiraling through a matter environment. We show that DF is not an additional force but a term at order $q^2\epsilon$ in a covariant multiparameter expansion of the MiSaTaQuWa force equation that vanishes identically in the absence of a particle-induced perturbation of the environment's matter fields, where $q$ and $\epsilon$ are respectively the mass ratio and environmental parameter. As a test of this identification, we evaluate it in weak and adiabatic field limit, solving the polar density-perturbation equation sourced by the particle's own order-$q$ vacuum metric perturbation. The reduction reproduces the Chandrasekhar-Ostriker drag force and predicts two features in strong gravity. An $\ell$-dependent splitting of the wake that only recombines into the Ostriker source in the weak-field limit, and a purely relativistic axial contribution with no Newtonian counterpart. This establishes dynamical friction as an intrinsic piece of self-force theory, extending to strong field, generic orbits, and generic environments, with direct consequences for extreme-mass-ratio inspiral waveform modeling for LISA.
\end{abstract}

\maketitle

\noindent \textbf{\textit{Introduction.}}--- Compact binaries with extreme mass ratios, $q\ll1$, are flagship sources for next-generation gravitational-wave (GW) detectors. Extreme- and intermediate-mass-ratio inspirals (EMRIs/IMRIs) will spend months to years in the LISA and TianQin bands, completing large number of orbital cycles as a stellar- or intermediate-mass compact object spirals into a $10^5$--$10^7\,M_\odot$ massive black hole~\cite{Berry:2019wgg,2017arXiv170200786A,TianQin:2015yph}. Their long-lived signals make EMRIs exceptional probes of both the strong-field gravity and the astrophysical environments, such as dark-matter overdensities, accretion disks, in which massive black holes are expected to reside \cite{Barausse:2020rsu,Bertone:2018krk}. Exploiting this information, however, requires waveform models with sub-radian phase accuracy throughout the inspiral. In vacuum spacetimes, the gravitational self-force (GSF) formalism provides such accuracy by expanding the Einstein equations in powers of mass ratio $q$. Only recently has the combination of first-order conservative and second-order dissipative GSF effects enabled complete first post-adiabatic (1PA) waveforms \cite{Barack:2018yvs,Hinderer:2008dm,Pound:2019lzj,Wardell:2021fyy,Warburton:2021kwk}.

Achieving comparable precision in astrophysical environments requires extending the self-force framework beyond vacuum GR and is currently lacking~\cite{Barausse:2014tra,Speri:2022upm,Cardoso:2019rou,Cardoso:2022whc,Speeney:2024mas}. Recently, Ref.~\cite{Datta:2025ruh} (DM) developed a framework for generic spherically symmetric environments by treating both the secondary and the surrounding matter as coupled perturbations in a multiparameter expansion in the mass ratio $q$ and the environmental parameter $\epsilon$ (Also see Ref.~\cite{Rahman:2025mip}). They follow the approach of Regge-Wheeler and Zerilli for computing perturbations \cite{Regge:1957td,Zerilli:1970wzz,Zerilli:1970se,Sago:2002fe, Martel:2003jj,Martel:2005ir}. Using the formalism, DM computes generic perturbation of the system and the emitted GW flux at the boundaries.

However, a compact object moving through matter also excites a wake that backreacts on its motion inducing dynamical friction (DF), which has not been explored in DM. Usually DF is modeled phenomenologically using Newtonian or relativistic prescriptions tailored to only some specific orbital configurations \cite{1943ApJ....97..255C,1971ApJ...165....1R, 1980ApJ...240...20R, 1989ApJ...336..313P, Syer:1994vr, Ostriker:1998fa,1999ApJ...522L..35S,Barausse:2007ph,Kim:2007zb,Kim:2008ab,Cashen:2016neh, Costa:2018gva, Katz:2019rgf, Traykova:2021dua,Vicente:2022ivh,Chiari:2022kas,Dyson:2024qrq,Dyson:2025dlj,Vicente:2025gsg,Dyson:2026ddd,Lui:2026uai}. Such approaches are not derived within a
covariant perturbative framework, making systematic extensions to generic orbits and strong fields difficult.
A generic, systematically improvable prescription for DF is therefore needed.
We show that within the multiparameter framework of Ref.~\cite{Datta:2025ruh}, DF emerges naturally from the matter-response sector of the GSF and hence can be extended to arbitrary orbits and geometry using DM's framework.

\noindent \textbf{\textit{Covariant environmental self-force.}}--- This identification follows directly from the structure of the MiSaTaQuWa equation~\cite{Mino:1996nk,Quinn:1996am}. A compact object inspiraling through matter, loses energy through both gravitational-wave emission and the dynamical response of the environment. While the former is described by the MiSaTaQuWa self-force equation ~\cite{Mino:2003yg}, the latter has traditionally been introduced as an independent drag force. Under DM's multiparameter framework, self-force expansion already contains a matter-coupled sector at the relevant order, and nothing beyond it is needed to generate the drag. 
Specifically, DF is not an additional force but the $\mathcal{O}(q^2\epsilon)$ matter-response component of the GSF, structurally alongside the vacuum radiation-reaction force appearing at order $q^2$.

In the small-mass-ratio limit, the spacetime metric
decomposes as $ g_{\mu\nu}= \bar g_{\mu\nu} + q h_{\mu\nu}$, where $\bar g_{\mu\nu}$ and $h_{\mu\nu}$ are respectively the background metric and metric perturbation, with the secondary of mass $\mu$ following an orbit obeying
the MiSaTaQuWa equation \cite{Mino:1996nk,Quinn:1996am,Detweiler:2002mi, Poisson:2003nc, Poisson:2011nh, Pound:2015tma,Barack:2018yvs},
\begin{equation}
\frac{Du^\mu}{d\tau} = -\frac{1}{2}\left(g^{\mu\nu}+u^\mu u^\nu\right)
\left(2 h^R_{\nu\lambda;\rho} - h^R_{\lambda\rho;\nu}\right) u^\lambda u^\rho,
\label{eq:misataquwa}
\end{equation}
where $u^{\mu}$ is its four velocity. The multiparameter expansion of DM separates the particle's mass ratio $q$ from the environmental parameter $\epsilon$ and decomposes all relevant quantities order by order,
\begin{equation}
g_{\mu\nu} = g^{(0,0)}_{\mu\nu} + \epsilon\, g^{(0,1)}_{\mu\nu}
+ q\, g^{(1,0)}_{\mu\nu} + q\epsilon\, g^{(1,1)}_{\mu\nu} + \dots ,
\label{eq:multiparam}
\end{equation}
where $(0,0)$ is the vacuum background set to the Schwarzschild metric, $(0,1)$ the static environmental deformation of the background, $(1,0)$ the standard vacuum Regge-Wheeler-Zerilli particle perturbation, and $(1,1)$ the coupled particle-environment response, identifying $\Bar{g}_{\mu\nu}=g^{(0,0)}_{\mu\nu} + \epsilon\, g^{(0,1)}_{\mu\nu}$ and $h_{\mu\nu}= g^{(1,0)}_{\mu\nu} + \epsilon\, g^{(1,1)}_{\mu\nu}$. For further details see supplemental material.

The vacuum GSF contributions enter the acceleration in Eq.~\eqref{eq:misataquwa} at $\mathcal{O}(q)$, and the environmental corrections at $\mathcal{O}(q\epsilon)$. Since the physical force is $F^i=\mu a^i$, these correspond to a force at one order higher in $q$, $\mathcal{O}(q^2)$ and $\mathcal{O}(q^2\epsilon)$ respectively. One of the obstacles in identifying the DF contribution is that the  GSF mixes two physically distinct effects. First, the particle's response to a static deformation of the background by the environment, and second, the particle's response to the environment's dynamical reaction to the particle itself. Only the latter constitutes drag in the ordinary sense and is a force sourced by the particle's own passage, absent otherwise.
For example, terms constructed from $h^{(1,0)}_{\lambda\rho;\nu}$ contracted with the $(0,1)$-order environmental static background metric or connection, modify the vacuum self-force and also contribute at $q^2\epsilon$-order. However, as they depend only on the particle's vacuum field and the static environmental deformations, they persist even in the complete absence of any particle-induced wake.
We take this as the defining criterion for DF within the self-force expansion.

\emph{A term in the $O(q^2\epsilon)$ self-force can be dynamical friction only if it vanishes identically when the environment's $(1,1)$-order matter perturbation is switched off.} This criterion separates static environmental corrections from the
medium's dynamical response, identifying the latter as the sector that contains DF.
Ref.~\cite{Barausse:2007ph} computed relativistic corrections to DF for specific fluid and orbital 
configurations using a linearized geodesic force law on a flat background. Here we identify DF, within the general multiparameter expansion, as the dissipative component of the matter-response sector of the GSF, which in principle applicable to generic orbits and background.
By the criterion above, the DF acceleration can only be sourced by the $q\epsilon$-order matter mediated acceleration,
\begin{equation}
-2\, a^\mu = \left(g^{(0,0)\mu\nu} + \mathcal{U}^{(0,0)\mu\nu}\right)
\left(2 h^{(1,1)R}_{\nu\lambda;\rho} - h^{(1,1)R}_{\lambda\rho;\nu}\right) \mathcal{U}^{(0,0)\lambda\rho} ,
\label{eq:sfforce}
\end{equation}
where $h^{(1,1)R}_{\mu\nu}$ is the regular field sourced by the point particle and the medium's response to the particle's gravitational field, and $\mathcal{U}^{\mu\nu}=u^{\mu} u^{\nu}$.

\noindent \textbf{\textit{Weak field reduction.}}--- In the weak field limit $g^{(0,0)}_{\mu\nu}$ identifies to Minkowski spacetime and $u^{(0,0)\mu}=\gamma v^{\mu}=\gamma\frac{dx^{\mu}}{dt}$, where $\gamma=\frac{dt}{d\tau}$, $t$ and $\tau$ respectively are the coordinate and proper time. For $\mathcal{U}^{(0,0)\mu\nu}=\gamma^2(1,0,0,v)\otimes(1,0,0,v)$ with the slow-velocity limit ($v\to 0,\,\gamma\to1$), Eq.~\eqref{eq:sfforce} reduces to the intuitive Newtonian-style force,
\begin{equation}
a^i \approx \tfrac{1}{2}\,\partial_i h^{(1,1)R}_{00}, \quad a^i_{\rm DF} \approx \tfrac{1}{2}\,\partial_i h^{(1,1)\delta\rho}_{00},
\label{eq:df_force}
\end{equation}
sourced by the $(1,1)$-order gravitational potential. The full acceleration $a^i$ receives contributions from all $(1,1)$-order source terms discussed in DM. We denote by $a^i_{\rm DF}$ the component obtained from Eq.~\eqref{eq:df_force} by restricting to the density-perturbation source (discussed later), with $R$ replaced by $\delta\rho$. This is the quantity compared below with the Chandrasekhar--Ostriker (CO) force.

\noindent \textbf{\textit{Density perturbation.}}--- We start from the density perturbation equation derived in DM for $\ell\geq 0$ polar sector,
\begin{align}\
(\partial_{r_\star}^2-c^{-2}_{r,{\ell m}}\partial_t^2+V^{\rho}_{\ell}
+\gamma_{1,\ell}\partial_{r_\star})\rho_{\ell m}^{(1,1)}=S_{\ell m}^{\rho}\ .\label{eq:masterrho} 
\end{align}
where $r_*$ is the tortoise coordinate, $V^{\rho}_{\ell}$ and $\gamma_{1,\ell}$ are some potentials and  $S_{\ell m}^{\rho}$ is the source term governed by $(1,0)$ order point particle motion and metric perturbations. The expressions are provided in \cite{DattaGitHub,SGREP_REPO}. To recover the CO expression, we solve Eq.~\eqref{eq:masterrho} for $\rho_{\ell m}^{(1,1)}$ in the weak-field limit, assuming a homogeneous isotropic medium ($p_t=p_r\equiv p$, $c_{r,\ell m}'=0$, $\rho'(r)=p'(r)=0$) with mode-independent sound speed $c_{r,\ell m}=c_s$. In the weak field limit $V^{\rho}_{\ell}\to -2(1+\Lambda)/r^2=-\ell(\ell+1)/r^2$ and $\gamma_{1,\ell} \to 2/r$. Multiplying Eq.~\eqref{eq:masterrho} with $Y_{\ell m}(\theta,\phi)$ and summing over $\ell$
and $m$ we find, 
\begin{equation}
\left(-\frac{\partial_t^2}{c_s^2}+\nabla^2\right) \rho^{(1,1)} = S_\rho = S^{(g)}_\rho + S^{(PP)}_\rho ,
\label{eq:waveeq}
\end{equation}
where $\nabla^2$ is the three dimensional flat space Laplacian operator. 
The source splits into two parts. A $(1,0)$ order point particle dependent contribution $S^{(PP)}_\rho$, sourced by the worldline-supported stress-energy tensor such as $\sum_{\ell m}\mathcal{A}^{0(1,0)}_{\ell m} Y_{\ell m}(\theta,\phi)$ and a metric component dependent piece $S^{(g)}_\rho$, sourced by the vacuum $(1,0)$-order metric functions. Since we are interested in reproducing the leading order CO expression, i.e. $\mathcal{O}(v^0)$ contribution, we ignore any point particle source term proportional to $u^r$, $u^{\theta}$, and $u^{\phi}$ from $S^{(PP)}_\rho$ as they contribute in the higher order. Similarly we ignore any time derivative of metric components like $\partial_tK^{(1,0)}_{\ell m}$ or any time derivative of $(1,0)$ order master function $\chi_{\ell m}^{(1,0)}$ from $S^{(g)}_\rho$.

The Green's function of the operator thus introduces the wake structure arising from causality~\cite{Ostriker:1998fa},
\begin{equation}
\rho^{(1,1)} =-\int d^3x' dt' \frac{\delta\!\left(t-t'-\dfrac{|\mathbf{x}-\mathbf{x}'|}{c_s}\right)}{4\pi|\mathbf{x}-\mathbf{x}'|}S_\rho,
    \label{eq:Flat_space_GF}
\end{equation}
where $(t,\mathbf{x})$ and $(t',\mathbf{x}')$ are respectively the field and source point, with boldface representing three vector.

\noindent \textbf{\textit{Explicit reduction.}}--- The perturbation equations in DM were handled according to the mode number $\ell$. Due to the structure of the perturbations, $\ell=0$, $\ell=1$, and $\ell>1$ must be tackled separately. For $\ell>1$ along with the  previous approximations in place, the density source simplifies considerably, leaving only surviving terms proportional to $\mathcal{A}^{0(1,0)}_{\ell m}$, $H_{0,\ell m}^{(1,0)}$, $K_{\ell m}^{(1,0)}$, and $\partial_r^nK_{\ell m}^{(1,0)}$ ($n\leq3$), fixed entirely by the vacuum field equation, where $h_{\theta\theta}^{(1,0)}\propto h_{\phi\phi}^{(1,0)} \propto K^{(1,0)}$ and $ h_{00}^{(1,0)} \propto H_{0}^{(1,0)}$ ($\ell,\,m$ indices are removed for undecomposed components). In this limit $K_{\ell m}^{(1,0)}$ satisfies a Poisson equation sourced by the particle's own monopole density,
\begin{equation}
\partial_r^2 K^{(1,0)}_{\ell m} + \frac{2}{r}\partial_r K^{(1,0)}_{\ell m} - \frac{\ell(\ell+1)}{r^2}K^{(1,0)}_{\ell m}
= -8\pi \mathcal{A}^{0(1,0)}_{\ell m}.
\label{eq:poisson}
\end{equation}
We use this equation to replace $\partial_r^2K_{\ell m}^{(1,0)}$ and $\partial_r^3K_{\ell m}^{(1,0)}$ in the source along with other Einstein equations to replace $H_{0,\ell m}^{(1,0)}$ in terms of $K_{\ell m}^{(1,0)}$. Substituting Eq.~\eqref{eq:poisson} back, the source can be simplified containing terms proportional to only $\mathcal{A}^{0(1,0)}_{\ell m}$, $K_{\ell m}^{(1,0)}$, and $\partial_rK_{\ell m}^{(1,0)}$.
The first term introduces $\delta(r-r_p)$, where the subscript $p$ labels the value at particle position. The latter two are smooth or at most jump at the particle position and hence cannot contribute to the exact point-source structure recovered below. They instead source a smoother correction to the wake. Dropping these and retaining only the delta-function term, the source for each $(\ell,m)$ mode becomes $-\frac{4 \pi  \gamma  \mu \kappa }{c_s^2 r^2}Y^*_{\ell m}(\Omega_p)  \delta (r-r_p)$, where enthalpy $\kappa=\rho+p$, and $\Omega$ is a shorthand for angular coordinates.

For $\ell=0$, using the $(1,0)$ order perturbation equations, the density equation takes a similar form as Eq.~\eqref{eq:waveeq}, where the source term $S^{(g)}_\rho\propto H_{2,00}^{(1,0)}$ and $S^{(PP)}_\rho\propto \mathcal{A}^{0(1,0)}_{00}$. Similarly for $\ell=1$, the source term $S^{(g)}_\rho\propto H_{0,1 m}^{(1,0)},\, H_{2,1m}^{(1,0)}$ and $S^{(PP)}_\rho\propto \mathcal{A}^{0(1,0)}_{1 m}$.
Post simplification, both $\ell=0,\,1$ source terms result in the same expression $-\frac{4 \pi  \gamma  \mu \kappa }{c_s^2 r^2}Y^*_{\ell m}(\Omega_p)  \delta (r-r_p)$. 
Using the identity $\sum_{\ell m} Y^*_{\ell m}(\Omega_p)Y_{\ell m}(\Omega) = \delta^2(\Omega-\Omega_p)$ and $\mathcal{A}^{0(1,0)}_{\ell m}$ expression from Ref.~\cite{Sago:2002fe}, the total source summed over all the modes become proportional to the three dimensional delta function,
\begin{equation}
    S_{\rho} = -\frac{4 \pi  \gamma  \mu   \kappa \delta^3(\mathbf{x}-\mathbf{x}') }{c_s^2}.
\end{equation}
Once expressed in terms of the fractional density contrast $\alpha\equiv\rho^{(1,1)}/\rho_0$ used by Ostriker, for a homogeneous background, Eq.~\eqref{eq:waveeq} then becomes 
\begin{equation}
    (\nabla^2-c_s^{-2}\partial_t^2)\alpha=-4\pi(\mu\kappa/\rho_0)\delta^3(\mathbf{x}-\mathbf{x}')/c_s^2,
\end{equation}
which reduces to $-4\pi\mu\delta^3(\mathbf{x}-\mathbf{x}')/c_s^2$ under $p=0,\,\gamma=1$ ($\kappa\to\rho_0$),
matching Ostriker's equation exactly, and hence provides exactly the same density wake solution.
From the (1,1) order Einstein equation, ignoring the time variations and solving for $ h_{00}^{(1,1)} \propto H_{0}^{(1,1)}$ we find,
\begin{equation}
\partial_r^2 H_{0,\ell m}^{(1,1)} + \frac{2}{r}\partial_r H_{0,\ell m}^{(1,1)} - \frac{\ell(\ell+1)}{r^2}H_{0,\ell m}^{(1,1)}
= S_{H_0}.
\label{eq:H011_poisson}
\end{equation}
Once multiplied with $Y_{\ell m}(\Omega)$ and summed over each $\ell$ and $m$ modes, the LHS becomes flat space Laplacian operator acting on $H_{0}^{(1,1)}$, reproducing the equations of Ref.~\cite{Ostriker:1998fa,Barausse:2007ph}. For $\ell>1$, $S_{H_0}$ in
Eq.~\eqref{eq:H011_poisson} contains four types of terms. Two proportional to $(1,0)$ order source or metric perturbations like $\mathcal{A}^{0(1,0)}_{\ell m}$ or $H_{0,\ell m}^{(1,0)}$ times background environmental quantities, and one proportional to $(1,1)$ order source like $\mathcal{A}^{0(1,1)}_{\ell m}$, independent of the $(1,1)$-order fluid perturbation. The fourth term is $-8\pi\rho^{(1,1)}_{\ell m}$ at leading order from matter perturbation. For $\ell=0$ and $1$, a similar structure prevails.
Linearity of Eq.~\eqref{eq:H011_poisson} decomposes $H_{0,\ell m}^{(1,1)}$ into four additive contributions, and hence four separate contributions to $a^i$ via Eq.~\eqref{eq:sfforce}, of which only the fourth survives the criterion. This sector requires the medium's dynamical response to the particle. 
In the polar sector this response takes the specific form of a causally retarded density wake. 
Using only the fourth source term in $S_{H_0}$ in Eq.~\eqref{eq:H011_poisson} provides the solution for the wake driven drag from $h^{(1,1)\delta\rho}_{00}=H_0^{(1,1)\delta\rho}=-2\Phi^{(1,1)\delta\rho}$ derived by Ostriker.
Since the source, governing operator, retarded Green's function, and (1,1) order metric equation all coincide with Ostriker, this potential agrees identically with the one in Ref.~\cite{Ostriker:1998fa} and the drag force obtained from it via Eq.~\eqref{eq:df_force} coincides with Ostriker's force integral.

This establishes that the identified matter-response sector reproduces the Chandrasekhar--Ostriker force in its appropriate limit.

\noindent \textbf{\textit{Beyond Chandrasekhar--Ostriker recovery}}--- The covariant formulation, however, predicts considerably more than this weak-field reduction.
Beyond the Chandrasekhar–Ostriker limit recovery, the formalism predicts two new relativistic effects: an $\ell$-dependent splitting of the wake and an axial contribution to the matter-mediated force. These effects originate directly from the exact strong-field structure of the leading order source $\mathcal{A}^{0(1,0)}_{\ell m}$. Retaining the full strong field dependence, the $\mathcal{A}^{0(1,0)}_{\ell m}$ dependent source becomes~\cite{Sago:2002fe, Nakano:2003he},
\begin{equation}
   -\frac{4 \pi  \gamma  f_p^2 \mu \kappa}{c_s^2}\,\,\frac{\delta\big(r-r_p(t)\big)}{r^2}\, \frac{  r (\Lambda  r+2)-2 }{r  (\Lambda  r+2)}\,Y^*_{\ell m}\big(\Omega(t)\big).
\label{eq:full_source}
\end{equation}
Unlike the weak-field limit, the source acquires a nontrivial radial
dependence.
It carries the factor $f_p^2=(1-2/r_p)^2$, driving the source to zero as $r_p\to2$. This is a relativistic
effect with no Newtonian counterpart, reflecting the strong gravity modification of the Newtonian potential gradient of the Ostriker mechanism.
Along with the source, in the strong gravity regime the operator acting on density perturbation in LHS of Eq.~\eqref{eq:masterrho} also modifies from flat space Laplacian due to $V_{\ell}^{\rho}$ and $\gamma_{1,\ell}$. Unlike the strong field effects in the source term these modifications affect the Green's function itself, modifying the causal wake structure in Eq.~\eqref{eq:Flat_space_GF}.

The first prediction in this strong-field limit is the splitting of the wake into mode-dependent contributions. Each $(\ell,m)$ source depends explicitly on $\ell$. Only in the $r\gg2$ limit, does
$\frac{  r (\Lambda  r+2)-2 }{r  (\Lambda  r+2)}\to 1$ becoming mode-independent, allowing the mode sum
to close via $\sum_{\ell m} Y^*_{\ell m}(\Omega_p)Y_{\ell m}(\Omega) = \delta^2(\Omega-\Omega_p)$ and recover the CO expression. Strong gravity therefore splits the wake into mode-dependent
contributions, implying mode-dependent constituents in full dynamical friction. The present reduction isolates only the leading-order contribution of this structure.

We neglected other point-particle delta-function sources in $S^{(PP)}_{\rho}$ such as, $\mathcal{B}^{0(1,0)}_{\ell m}$, $\mathcal{F}^{(1,0)}_{\ell m}$, $\mathcal{A}^{1(1,0)}_{\ell m}$, $\mathcal{A}^{(1,0)}_{\ell m}$, and $\mathcal{B}^{(1,0)}_{\ell m}$, as they carry at least an extra time derivative. They contribute to the wake similarly, with mode splitting, but depend on the detailed orbital configuration. For a circular orbit $\mathcal{A}^{1(1,0)}_{\ell m}=\mathcal{A}^{(1,0)}_{\ell m}=\mathcal{B}^{(1,0)}_{\ell m}=0$ ($\propto
u^r$), though nonzero for eccentric orbits, while $\mathcal{F}^{(1,0)}_{\ell m}=0$ ($\propto u^{\theta},u^\phi$), for radial infall. Each depends differently on $f_p$, varying in importance in the
strong-field limit.
We similarly ignored source terms built directly from the $(1,0)$-order master function and its derivatives that do not produce delta functions, contributing at higher PN order which should modify the DF as well.

The second genuinely new relativistic effect
appears outside of the polar sector altogether. Eq.~\eqref{eq:sfforce} contains a contribution from the axial gravitational sector, found here for the first time. The vacuum axial metric perturbation $h_{0,\ell m}^{(1,0)}$ ($h^{(1,1)}_{t\phi}$) generates, to leading order, a radial force,
\begin{equation}
\label{eq:axial sf}
a^r_{\rm axial} \approx
\gamma^2 v^\phi\,\partial_r h^{(1,1)R}_{t\phi},
\end{equation}
with no analog in any Newtonian treatment. Its precise role within the matter-response sector is
naturally addressed through the flux-balance formulation discussed below.

In Eq.~\eqref{eq:df_force} we retained only the $\mathcal{U}^{(0,0)tt}$ piece of the four-velocity contraction in Eq.~\eqref{eq:sfforce}. The full contraction also sources a $\gamma^2v^2$ piece via $\mathcal{U}^{(0,0)\phi\phi}$, together with the axial cross term already isolated in Eq.~\eqref{eq:axial sf}. Both are sourced by the same matter perturbation used to recover the Chandrasekhar-Ostriker force above. Eq.~\eqref{eq:df_force} therefore generalizes to $a^i_{\rm DF}\approx\tfrac12\gamma^2(1+v^2)\,\partial_i
h^{(1,1)\delta\rho}_{00}$, reproducing the relativistic velocity prefactor $\gamma^2(1+v^2)$ of
Ref.~\cite{Barausse:2007ph} directly from the covariant self-force. Completing the identification to
the full $\gamma^3(1+v^2)^2$ prefactor requires the source itself ($\mathcal{A}^{0(1,0)}_{\ell
m}\propto\mu\gamma$) and the neglected sources such as $\mathcal{G}^{0(1,0)}_{\ell m}, \mathcal{F}^{0(1,0)}_{\ell m}$. Completing these polar and axial analysis requires including these additional source terms together with the flux-balance formulation and lies beyond the scope of the present Letter.

\noindent \textbf{\textit{Physical interpretation.}}--- These results also admit a simple physical interpretation. The particle excites environmental degrees of freedom that generate a retarded gravitational response, which acts back on the particle. Radiation reaction and dynamical friction are therefore not distinct mechanisms. Both are retarded responses to the particle's own passage. They differ only in which degrees of freedom carry the dissipated energy and angular momentum away, gravitational radiation propagating to infinity and the horizon, or matter perturbations excited locally in the surrounding environment. Both channels operate simultaneously whenever an environment is present. DF is simply the local, matter-mediated counterpart to the familiar radiative one.

This viewpoint naturally leads to a complementary flux -balance formulation, the local, near-zone analogue of the asymptotic GW flux balance already used in vacuum self-force theory. There, the dissipative effects can be obtained equivalently in two ways. Either by computing the MiSaTaQuWa force directly, or by balancing the orbital energy and angular momentum lost by the particle against the flux radiated away. Since a drag force is dissipative by definition, the particle must lose energy and angular momentum whenever it experiences one, and this loss can likewise be tracked through a flux balance. In vacuum, that balance is against the asymptotic GW fluxes and here, the analogous balance is instead against the energy and angular momentum injected into the environment. This channel has not been explored in DM, which accounts only for the GW flux radiations at infinity and horizon. A full flux-balance derivation of $a^\mu_{\rm DF}$, carried out in exact analogy to these vacuum GSF methods, thus can provide an independent, and potentially a computationally cheaper path to DF.
Such a balance additionally offers a natural route to separating the dynamical-friction component from the full matter-response self-force, thereby identifying any conservative remainder and will be presented elsewhere.

\noindent \textbf{\textit{Conclusions.}}--- Taken together, these results establish DF as a subset of the environmental gravitational self-force. The construction reproduces the Chandrasekhar--Ostriker drag force in the weak-field limit while predicting two genuinely relativistic effects beyond it, an $\ell$-dependent wake and an axial matter-mediated force.
The density source itself further exhibits an exact relativistic redshift factor $(1-2/r_p)^2$ arising as strong field correction. 
Conceptually, environmental dissipation and GW radiation reaction are manifestations of the same retarded self-interaction, differing only in whether the response is carried by matter perturbations or GWs.
Denoting the complete wake-mediated drag identified above by $F_{\rm DF}^{\rm full}$, and the entire matter-response sector by $F_{\rm Matter}^{\rm full}$, the resulting picture organizes into a hierarchy,
\begin{equation}
 F_{\rm CO}\subset F_{\rm Barausse}
\subset F_{\rm DF}^{\rm full}\subset F_{\rm Matter}^{\rm full} \subset F_{\rm SF},
\label{eq:hierarchy}
\end{equation}
in which each successively more relativistic drag prescription is recovered as a controlled approximation to the covariant self-force, and $F_{\rm DF}^{\rm full}$ additionally includes the full generic-orbit and strong gravity corrections.
$F_{\rm Matter}^{\rm full}$, identified by the vanishing criterion above, is a strict superset of
$F_{\rm DF}^{\rm full}$. It requires the environment's dynamical response but need not be dissipative, whose classification is left open pending the flux-balance analysis discussed above. This separation is possible because the multiparameter expansion treats the static environmental background, $g_{\mu\nu}^{(0,1)}$, independently of the dynamical response.

Unlike prior relativistic treatments, tied to a fixed orbit family and a flat or patched background, the criterion and field equations used here are orbit- and profile-generic by construction. The density potentials depend only on the background, not on the source, so that extensions to eccentric or inclined orbits, anisotropic or inhomogeneous media, and higher post-Newtonian order require only new source or equation of state input, not a new derivation.
The construction assumes the secondary retains a fixed mass throughout the inspiral. A secondary black hole however will accrete matter from the environment, making its mass dynamical. As a result the force may receive an additional contribution that has not been explored here. Whether this accretion induced contribution is already captured by the criterion above, or requires extending the underlying force law to a dynamical-mass secondary remains to be studied.

Beyond providing a first-principles derivation, this work changes the theoretical status of dynamical friction.
It emerges naturally as a sector of the GSF, providing a systematic route toward precision EMRI modelling in generic astrophysical environments.

\noindent \textbf{\textit{Acknowledgments.}}--- The author thanks Andrea Maselli for suggesting the project and for providing meaningful inputs.
The author also thanks Enrico Barausse for reading an earlier version and providing useful inputs.
The author acknowledges financial support from MUR, PNRR - Missione~4 - Componente~2 -
Investimento~1.2 - finanziato dall'Unione europea - NextGenerationEU (cod.\ id.\:
SOE2024\_0000167, CUP: D13C25000660001).
The author acknowledges CINECA for providing high-performance computing resources and support through the ISCRA initiative under project HP10CU7X29.

\newpage

\appendix
\section{The multiparameter expansion}
We follow Ref.~\cite{Sago:2002fe, Datta:2025ruh} to decompose the metric and point particle energy momentum tensor source. The background in Schwarzschild coordinates $x^\mu=(t,r,\theta,\phi)$ is
\begin{equation}
g^{(0,0)}_{\mu\nu}=\mathrm{diag}\!\left(-f,\,f^{-1},\,r^2,\,r^2\sin^2\theta\right),
\qquad f=1-\frac{2M}{r}\, .
\end{equation}
At order \((0,1)\) the metric components are,
\begin{equation}
g_{\mu\nu}^{(0,1)} = \text{diag}\left(-f H, \frac{2m}{r f^{2}}, 0,0\right)\ ,
\end{equation}
that is determined by the corresponding density and pressures.
Due to the symmetry of the background, metric perturbations 
can be separated into the usual families of axial ($A$) and polar ($P$) components as~\cite{Regge:1957td,Zerilli:1970wzz,Zerilli:1970se}:
\begin{equation}
\delta g_{\mu\nu}(x^\alpha) = \delta g_{\mu\nu}^{A}(x^\alpha) +\delta  g_{\mu\nu}^{P}(x^\alpha)\ .
\end{equation}

\begin{widetext}
\begin{align}
\delta g_{\mu\nu}^{A} =&  \sum_{\ell,m} \frac{\sqrt{2\lambda}}{r} \left[ ih_{1,\ell m}(t,r) {\bf c}_{\ell m}(\theta,\phi) -h_{0,\ell m}(t,r) {\bf c}^0{}_{\ell m}(\theta,\phi)   
+\frac{\sqrt{\Lambda}}{r}h_{2,\ell m}(t,r){\bf d}_{\ell m}(\theta,\phi)
    \right]\ ,\label{eq:axial10}\\
\delta g_{\mu\nu}^{P} =& \sum_{\ell,m}\Big[-\bar{g}_{tt} H_{0,\ell m} (t,r) {\bf a}^0{}_{\ell m}(\theta,\phi) - i \sqrt{2} H_{1,\ell m} (t,r) {\bf a}^1{}_{\ell m}(\theta,\phi)
- \frac{i}{r}\sqrt{2\lambda} \eta_{0,\ell m} (t,r) {\bf b}^0{}_{\ell m}(\theta,\phi)+\frac{\sqrt{2\lambda}}{r} \eta_{1,\ell m} (t,r) {\bf b}{}_{\ell m}(\theta,\phi) \nonumber\\
&+\bar{g}_{rr}H_{2,\ell m} (t,r) {\bf a}{}_{\ell m}(\theta,\phi)
+\sqrt{\Lambda \lambda}G_{\ell m} (t,r)  {\bf f}_{\ell m}(\theta,\phi)+ \left(\sqrt{2}  K_{\ell m} (t,r)-\frac{\lambda}{\sqrt{2}}G_{\ell m} (t,r)\right) {\bf g}_{\ell m}(\theta,\phi)
\Big]\ ,\label{eq:polar10}
\end{align}
\end{widetext}
where $\lambda=\ell(\ell+1)$, $\Lambda=(\ell+2)(\ell-1)/2$, and 
the sum over the multipolar indices $(\ell,m)$ runs from $\ell=0,\ldots,\infty$ and $m=-\ell,\ldots,\ell$. The ten basis components $\{{\bf c}^{\ell m}_{\mu\nu},{\bf c}^{0\ell m}_{\mu\nu}\ldots {\bf g}^{\ell m}_{\mu\nu}\}$ depend on the spherical harmonics $Y_{\ell m}(\theta,\phi)$ and their derivatives~\cite{Sago:2002fe}. Among the ten unknown functions $\{h_{1 \ell m} \ldots K_{\ell m}\}$, the 
axial term $h_{2\ell m}$ and the three polar components  $\{\eta_{0\ell m},\eta_{1\ell m},G_{\ell m}\}$ are set to zero using the Regge-Wheeler-Zerilli gauge. To decouple the vacuum and environmental perturbation we write $\delta g_{\mu\nu}=h_{\mu\nu}= g^{(1,0)}_{\mu\nu} + \epsilon\, g^{(1,1)}_{\mu\nu}$. Consequently, all metric components decomposes in multiparameter hierarchy, such as $H_{0,\ell m} (t,r)=H_{0,\ell m}^{(1,0)} (t,r) + \epsilon H_{0,\ell m}^{(1,1)} (t,r)$.
Similarly to the metric perturbations, the particle stress-energy tensor is also decomposed in the basis of tensor harmonics,
\begin{widetext}
\begin{align}
\label{eq:PP-EMT10}
    T^{p}_{\mu\nu}{}^{(1,0)} =\sum_{\ell,m}\Big[&\mathcal{A}^{0(1,0)}_{\ell m} {\bf a}^0{}_{\ell m}(\theta,\phi) +\mathcal{A}^{1(1,0)}_{\ell m} {\bf a}^1{}_{\ell m}(\theta,\phi) + \mathcal{A}^{(1,0)}_{\ell m} {\bf a}{}_{\ell m}(\theta,\phi) + \mathcal{B}^{0(1,0)}_{\ell m}{\bf b}^0{}_{\ell m}(\theta,\phi) + \mathcal{B}^{(1,0)}_{\ell m}{\bf b}{}_{\ell m}(\theta,\phi)\nonumber \\
    &+\mathcal{Q}^{(1,0)}_{\ell m} {\bf c}_{\ell m}(\theta,\phi) +\mathcal{Q}^{0(1,0)}_{\ell m} {\bf c}^0{}_{\ell m}(\theta,\phi) +  \mathcal{D}^{(1,0)}_{\ell m} {\bf d}_{\ell m}(\theta,\phi)+  \mathcal{G}^{(1,0)}_{\ell m} {\bf g}_{\ell m}(\theta,\phi) +  \mathcal{F}^{(1,0)}_{\ell m} {\bf f}_{\ell m}(\theta,\phi) \Big]\ .
\end{align}
\end{widetext}
The specific form of the coefficients $\{\mathcal{A}^{0(1,0)}_{\ell m},\ldots \mathcal{F}^{(1,0)}_{\ell m}\}$ depends on the 
secondary orbital configurations. The form of $T^p_{\mu\nu}{}^{(1,1)}$ can be constructed using the 
same ansatz of  Eqs.~\eqref{eq:PP-EMT10}, and replacing the functions with the correct order of the expansion, e.g. $\mathcal{Q}^{(1,0)}_{\ell m}\rightarrow \mathcal{Q}^{(1,1)}_{\ell m}$ (see Ref.~\cite{Sago:2002fe,Datta:2025ruh} for further 
details).

The $(1,1)$ contributions to the matter stress-energy 
tensor depend on the energy density and the pressure perturbations:
\begin{eqnarray}
    \rho =& \rho(r) + \rho^{(1,1)}(t,r,\theta,\phi)\ ,\\
    p=& p_r(r) + p_r^{(1,1)} (t,r,\theta,\phi)\ ,\\
    p_t =& p_t(r) + p_t^{(1,1)} (t,r,\theta,\phi)\ .
\end{eqnarray}

By exploiting the symmetry of the background to separate angular and time-radial variables, we expand fluid variables in terms of standard spherical harmonics,
\begin{align}
    \rho^{(1,1)}=& \sum_{\ell, m} \rho^{(1,1)}_{\ell m}(t,r) Y_{\ell m}(\theta,\phi)\ , \\
    p_r^{(1,1)} =&   \sum_{\ell, m} p^{(1,1)}_{r,\ell m}(t,r) Y_{\ell m}(\theta,\phi)
     ,\\
    p_t^{(1,1)}=&  \sum_{\ell, m} p^{(1,1)}_{t,\ell m}(t,r) Y_{\ell m}(\theta,\phi)\ .
\end{align}
Moreover, pressure and density perturbations are 
linked by an equation of state, such that,
\begin{equation}
p^{(1,1)}_{r,\ell m} = c^2_{r,{\ell m}}(r) \rho^{(1,1)}_{\ell m}\quad \ , \quad p^{(1,1)}_{t,\ell m} = c^2_{t,{\ell m}}(r) \rho^{(1,1)}_{\ell m}\ ,\label{eq:csdef}
\end{equation}
where the tangential $(c^2_{t,{\ell m}})$ and the radial $(c^2_{r,{\ell m}})$ sound speeds are in general not constant. 

Perturbations of the fluid velocity $u^\mu$, $u^{\mu(1,0)}$, can be written in terms of vector harmonics \cite{PhysRevD.46.4289, Datta:2025ruh} and expressed as,
\begin{align}
    u^{t(1,0)}&= \frac{1}{2\sqrt{f}} \sum_{\ell,m}H^{(1,0)}_{0,\ell m}(t,r) Y_{\ell m}(\theta,\phi)\ , \\
    u^{r(1,0)} &= \frac{f^{3/2}}{4\pi  }\sum_{\ell,m}W^{(1,0)}_{\ell m}(t,r) Y_{\ell m}(\theta,\phi)\ , \\
    u^{\theta(1,0)} &= \frac{\sqrt{f}}{4\pi  r^2}\sum_{\ell,m}\Big[V^{(1,0)}_{\ell m}(t,r) \partial_{\theta}\nonumber\\
    &\quad\quad\quad\quad\quad- \frac{U^{(1,0)}_{\ell m}(t,r)}{\sin \theta} \partial_{\phi} \Big]Y_{\ell m}(\theta,\phi)\ ,\\
    u^{\phi(1,0)} &= \frac{\sqrt{f}}{4\pi  r^2\sin^2\theta}\sum_{\ell,m}\Big[V^{(1,0)}_{\ell m}(t,r) \partial_{\phi}  \nonumber\\
    &\quad\quad\quad\quad\quad + U^{(1,0)}_{\ell m}(t,r)\sin \theta\partial_{\theta} \Big]Y_{\ell m}(\theta,\phi)\ .
\end{align}

\bibliography{references}

\end{document}